\documentclass{emulateapj}

\newcommand{\Mo}{$\mathrm{M_\odot}$}

\shorttitle{Period changes in six contact binaries} \shortauthors{Zasche et al.}

\begin{document}

\title{Period changes in six contact binaries: WZ~And, V803~Aql, DF~Hya, PY~Lyr, FZ~Ori, and AH~Tau}

\author{P. Zasche\footnote{eMail to: zasche@sirrah.troja.mff.cuni.cz}}
\affil{Astronomical Institute, Faculty of Mathematics and Physics, Charles University of Prague,
   CZ-180 00 Praha 8, V Hole\v{s}ovi\v{c}k\'ach~2, Czech Republic}

\author{A. Liakos and P. Niarchos}
\affil{Department of Astrophysics, Astronomy and Mechanics, Faculty of Physics,
   National \& Kapodistrian University of Athens, Athens, Greece}

\author{M. Wolf}
\affil{Astronomical Institute, Faculty of Mathematics and Physics, Charles University of Prague,
   CZ-180 00 Praha 8, V Hole\v{s}ovi\v{c}k\'ach~2, Czech Republic}

\author{V. Manimanis}
\affil{Department of Astrophysics, Astronomy and Mechanics, Faculty of Physics,
   National \& Kapodistrian University of Athens, Athens, Greece}

\author{K. Gazeas}
\affil{Harvard-Smithsonian Center for Astrophysics, 60 Garden Street, Cambridge, MA 02138, USA}

\begin{abstract}
Six contact binaries lacking a period analysis have been chosen to search for the presence of a
third body. The $O-C$ diagrams of these binaries were analyzed with the least-squares method by
using all available times of minima. Ten new minima times, obtained from our observations, were
included in the present research. The Light-Time Effect was adopted for the first time as the main
cause for the detailed description of the long-term period changes. Third bodies were found with
orbital periods from 49 up to 100~years, and eccentricities from 0.0 to 0.56 for the selected
binaries. In one case (WZ~And) a fourth-body LITE variation was also applied. The mass functions
and the minimal masses of such bodies were also calculated and a possible angular separation and
magnitude differences were discussed for a prospective interferometric discovery of these bodies.
\end{abstract}


\section{Introduction}

Eclipsing Binaries (hereafter EBs) are excellent objects for
determining the physical properties of stars and detecting
additional components in them. The long-time behavior of the
period of an EB could reveal the presence of another component
orbiting with the EB around the common center of mass. Photometric
observations of EBs sometimes cover more than a century, therefore
it is possible to detect the third bodies with a similar period.

The motion around the barycenter causes apparent changes of the
observed binary's period with a period corresponding to the
orbital one of the third body, called the LIght-Time Effect (or
'light-travel time', hereafter LITE). \cite{Irwin1959} improved
the method developed by \cite{Woltjer1922} for analyzing the
long-term variation of the times of minima caused by a third body
orbiting the eclipsing pair. Useful comments and limitations were
discussed by \cite{FCH73} and by \cite{Mayer1990}. Nowadays there
are more than one hundred EBs showing LITE, where the effect is
certainly presented or supposed (see e.g. \cite{BorkovitsHegedus},
\cite{Albayrak1999}, \cite{Wolf2004}, \cite{Hoffman2006}, etc.).
See the catalogue of the $O-C$ diagrams by \cite{Kreiner2001},
where the apparent orbital period changes in many EBs are
presented. The look of $O-C$ diagrams in the present study was
adopted to be the same as in this catalogue. In our figures
\ref{FigWZAnd1} to \ref{FigAHTau} the full circles represent the
primary and the open circles the secondary times of minima, the
bigger the point, the bigger the weight. For the limitations and
consequences of the $O-C$ diagram analysis, see e.g.
\cite{Sterken2005}.

The computation of the parameters of the third-body orbit is a classical inverse problem with 5
parameters to be found -- $p_3$, $T_0$, $A$, $\omega$, $e_3$, which indicate the period of the
third body, the periastron passage, the semi-amplitude of the light-time effect, the argument of
periastron and the eccentricity, respectively (for a detailed description see e.g.
\citealt{Mayer1990}). The ephemerides for the individual systems ($JD_0$ and $P$ for the linear one
and $q$ for the quadratic one) have to be calculated together with the parameters of LITE. The mass
function $f(M_3)$ and the minimal mass of the third component $M_{3,min} = M_3 \cdot \sin i_3$ (for
$i_3 = 90^\circ$) could be computed from this set of parameters. The weights assigned to individual
observations were used as following: $w=1$ for visual observations, 3-5 for photographic and 10 for
CCD and photoelectric observations. The computing code, used in the present work, could be
downloaded via the webpages of the
author\footnote{\href{http://sirrah.troja.mff.cuni.cz/~zasche/}{http://sirrah.troja.mff.cuni.cz/$\sim$zasche/}}.

The present study was carried out following a similar analysis of period changes in Algol-type
binaries (made by \citealt{ZLWN2008}). All the systems selected for this paper are contact EBs, of
W~UMa or $\beta$~Lyr type, with components of similar spectral types (ranging from F to K). Except
for a few new observations given in Table \ref{Minima}, all the times of minima used in this paper
were collected from the published literature and from minima databases available in the internet.

According to a recent paper on the period changes in Algols by \cite{Hoffman2006}, there could be a
connection between the spectral type of the secondary component and the nature of the period
changes. Systems with spectral types of secondaries later than F5 show $O-C$ variations, which
could be caused by magnetic activity cycles and convective envelopes. This effect was discussed by
\cite{Hall1989}, \cite{Applegate1992}, \cite{Lanza1998}, etc. The role of the magnetic cycles in
the period changes is discussed below, but, due to lack of information about the systems, such an
analysis is a difficult task. For some of the systems studied in this paper the spectral types of
the secondaries are known with a low confidence level, light-curve analysis is missing and
spectroscopy has never been done.

\begin{table}[h!]
\caption{New times of minima based on CCD observations (Kwee-van Woerden (\citeyear{Kwee}) method
was used).} \label{Minima} \centering
\begin{tabular}{c c c c c }
\hline
 Star   & HJD-2400000 & Error   & Type & Filter   \\
\hline
 DF Hya & 54189.34526 & 0.00018 &  II  &    R     \\
 DF Hya & 54210.33806 & 0.00005 &  I   &    R     \\
 FZ Ori & 54099.45835 & 0.00019 &  II  &    R     \\
 FZ Ori & 54102.45888 & 0.00006 &  I   &    R     \\
 FZ Ori & 54109.45926 & 0.00008 &  II  &    R     \\
 FZ~Ori & 54535.24562 & 0.00012 &  I   &    VR    \\
 AH Tau & 54099.32718 & 0.00006 &  II  &    R     \\
 AH Tau & 54115.29518 & 0.00004 &  II  &    R     \\
 PY Lyr & 54564.55117 & 0.00021 &  I   &    R     \\
 PY Lyr & 54592.51913 & 0.00017 &  II  &    R     \\
\hline
\end{tabular}
\end{table}

\section{Observations}

The new measurements were secured at the Athens University observatory, which is situated at Athens
University campus, Athens, Greece. The 40-cm telescope is equipped with the SBIG ST-8XMEI CCD
camera and the Bessell UBVRI photometric filters. All of the observations were obtained from
December of 2006 to April 2008. The exposure times depend on the individual systems and observing
conditions, ranging from 25 to 80 seconds.

\section{Analysis of individual systems}

\subsection{WZ~And}

The first analyzed system is the eclipsing binary WZ~And
(GSC~02799-01250). This $\beta$~Lyrae system is about 11.6 mag
bright in $V$ filter \citep{1971GCVS} and its spectrum was
classified as F5 + G3 (according to \citealt{1982PASP}). It was
discovered to be a variable by Miss Leavitt, see
\cite{Leavitt1923}, and its eclipsing nature was confirmed by
Zessewitsch (\citeyear{1925AN}). The most recent detailed analysis
of the light curve of this system was made by
\cite{WZAnd2006NewA}, which indicates that the system is in a
shallow contact and has the photometric mass ratio about 1.
Regrettably, the radial velocity curve has not been obtained so
far.

\begin{figure}[h]
 \plotone{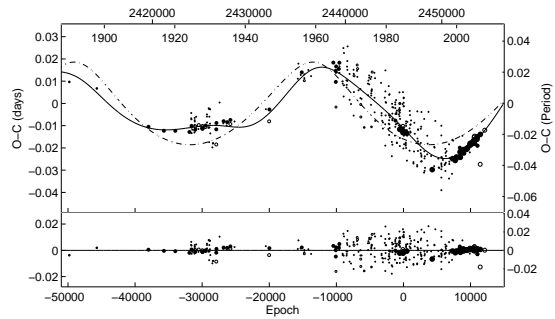}
 \caption{The $O\!-\!C$ diagram of WZ~And. The individual observations are shown as dots
 (primary minima) and open circles (secondary minima), the small ones for visual and the
 large ones for CCD and photoelectric observations, the bigger the symbol, the bigger the
 weight. The dash-dotted curve represents the predicted LITE variation caused by the third
 body, while the solid line represents the final fit (the third and the fourth body, see the
 text for details). The residuals after the fit are in the bottom part of the figure. For
 the variation caused by the fourth body only see Fig.\ref{FigWZAnd2}.
 \label{FigWZAnd1}}
\end{figure}

The first minima observations are more than a century old. Altogether there are about 400 minima,
but for the current analysis 37 of them were neglected due to their large scatter. The previous
period analysis of this system was performed by \cite{WZAnd2006NewA}, who suggested a few period
jumps and mass transfer between the two components. They were not able to fit a sinusoidal curve to
those data, because the period variation is more complicated. Also the previous analyses by
\cite{1991PASA} and \cite{1982PASP} presented a quadratic ephemeris to describe the variation in
the $O-C$ diagram.

\begin{figure}[b]
 \plotone{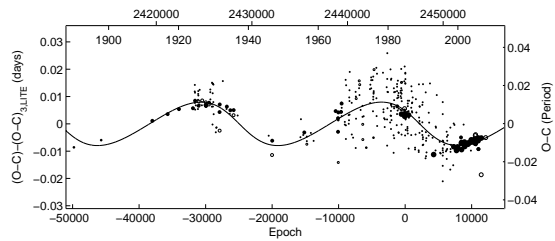}
 \caption{The $O-C$ diagram of WZ~And after subtraction of the variation caused by the third
 body. The solid line represents the predicted variation caused by a fourth body only.}
 \label{FigWZAnd2}
\end{figure}

With the new larger data set, one is able to identify a double variation in the $O-C$ diagram,
where the first one is caused by a third and the second one by a fourth body in the system (see
Fig.\ref{FigWZAnd1} for the plot of the $O-C$ diagram with the final fit). The $O-C$ variation, due
only to the fourth body is given in Fig.\ref{FigWZAnd2}. These variations have a  period of about
70 and 50 years respectively and their parameters are given in Table \ref{TableBig1}.

The total mass of the binary is $M_{12} = 2.27$~\Mo~(according to \citealt{1948Cook}). Using this
mass, one could calculate the mass function and the minimal masses of the third and also the fourth
component (see Table \ref{TableBig1}). These masses, assuming that the additional components belong
to the MS, indicate their spectral types should be about M3 and M5, respectively. Their
contribution to the total light is negligible (below 1\%), but in the red part of the spectrum
could be detectable. The distance of this system is about 440~pc \citep{1994Shaw}, so the predicted
angular separation and the magnitude difference between the additional components and the EB can
also be estimated. The angular separation results in 52 and 45~mas, respectively, which is well
within the limits for a detection by modern stellar interferometers. On the other hand, due to
their relatively low masses (and therefore also their luminosities), the magnitude differences
between them and the EB result in high values about 5~mag for the third and 6~mag for the fourth
component. Such large magnitude differences make extremely difficult the discovery of such distant
components.

If there are more than 3 components in the system, then a question arises not only about the
dynamical stability of such a system, but also about the suitability of the period analysis to find
such bodies. The method of LITE itself was derived on the basis of von Zeipel's method for the
three-body problem, where the assumption that the third body is far away from the eclipsing pair
does play the main role. This condition in the present case is not satisfied, so it is problematic
to judge whether this method could be used for such an analysis.

\begin{table*}
\small \caption{The final results (part 1): The parameters of the LITE orbits derived from the
$O-C$ analysis. The table is divided into two parts. In the first one, the computed parameters are
presented, while in the second one the derived quantities are given (the mass $M_{12} = M_1 + M_2$
and the distance $d$ are taken from the literature).} \label{TableBig1} \centering \scalebox{0.8}{
\begin{tabular}{c c c c c }
\hline
 Parameter         &   \multicolumn{2}{c}{WZ~And}                     &        V803~Aql          &           DF~Hya     \\
                   &3$^{\mathrm{rd}}$ body   & 4$^{\mathrm{rd}}$ body &                          &                       \\
 \hline
 $JD_0$ $[$HJD$]$  &  \multicolumn{2}{c}{$2446025.734 \pm 0.002$}      &$2447684.785 \pm 0.003$  & $2445021.531 \pm 0.004$ \\
 $P$ $[$day$]$     &  \multicolumn{2}{c}{$0.6956612 \pm 0.0000001$}    &$0.2634232 \pm 0.0000001$& $0.3306022 \pm 0.0000003$\\
 $p_3$ $[$yr$]$    &    $67.8 \pm 4.1$        &     $51.7 \pm 4.3$     &     $74.6 \pm 3.1$      &      $86.3 \pm 14.2$      \\
 $T_0$ $[$HJD$]$   &   $2437000 \pm 2400$     &   $2447500 \pm 2700$   &          --             &   $2430920 \pm 8000$     \\
 $\omega$ $[$deg$]$&     $98.4 \pm 37.5$      &      $179 \pm 58$      &          --             &       $165 \pm 104$      \\
 $e$               &    $0.28 \pm 0.18$       &     $0.23 \pm 0.22$    &   $0.000 \pm 0.110$     &     $0.162 \pm 0.096$    \\
 $A$ $[$day$]$     &  $0.0186 \pm 0.0022$     &   $0.0080 \pm 0.0024$  &  $0.0336 \pm 0.0023$    &     $0.054 \pm 0.009$    \\
 \hline
 $M_{12}$ [\Mo]    &      \multicolumn{2}{c}{$2.27$}                   &        $1.5$            &         $1.45$           \\
 $d$ [pc]          &      \multicolumn{2}{c}{$440$}                    &        $300$            &         $190$            \\
 $f(M_3)$ [\Mo]    &  $0.0073 \pm 0.0005$     &   $0.0011 \pm 0.0003$  &  $0.0353 \pm 0.0004$    &   $0.114 \pm 0.003$      \\
 $M_{3,min}$ [\Mo] &   $0.37 \pm 0.03$        &     $0.21 \pm 0.04$    &   $0.52 \pm 0.02$       &    $0.84 \pm 0.02$       \\
 $a$ [mas]         &        $52$              &          $45$          &        $75$             &         $150$            \\
 \hline
\end{tabular}}
\end{table*}

\subsection{V803~Aql}

The EB system V803~Aql is a neglected W~UMa-type system, with an orbital period about 0.3~days and
a depth of both primary and secondary minima about 0.8~mag. It is relatively faint binary, only
about 14~mag in $V$ filter and the spectral types of the components were found to range between K3
to K5 (see \citealt{Samec1993PASP}). The only detailed analysis of its light curve was made by
\cite{Samec1993PASP}, who derived that both components are very similar to each other, but also
proposed a possible explanation of its period changes due to the mass loss from the system caused
by a stellar wind.

\begin{figure}[b]
 \plotone{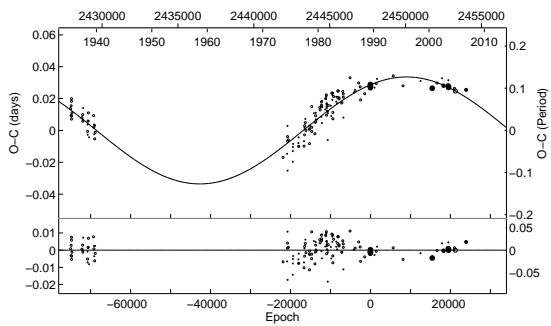}
 \caption{The $O-C$ diagram of V803~Aql (upper part), where the solid line represents the
 theoretical LITE variation caused by a 3$^\mathrm{rd}$ body and the $O-C$ residuals obtained
 after the subtraction of LITE (lower part). For the explanation of symbols see Fig.\ref{FigWZAnd1}.}
 \label{FigV803Aql}
\end{figure}

The new set of times of minima comprises 150 data points, from which 4 visual ones were neglected
due to their large scatter. The six minima times from \cite{Samec1993PASP} were corrected for their
heliocentric correction and fit better the theoretical curve of the LITE variation (see
Fig.\ref{FigV803Aql}). Such a variation is caused by a third body orbiting about the EB pair on its
75~yr circular orbit. The parameters of the orbit are given in Table \ref{TableBig2}, and these
values yielded a third-body mass about 0.5~\Mo. With the assumption that the third component is a
main sequence star, one gets its spectral type about K8. Such a body could be detectable in the
light curve analysis, as well as in the spectra of the system. Regrettably, no spectral analysis
has been performed so far and \cite{Samec1993PASP} did not include the third light parameter in
their light curve analysis.

Despite the fact that the distance of the system is not known, one can estimate a photometric
parallax of V803~Aql on the basis of its spectral type and luminosity. The distance is therefore
about 300~pc, from which one could estimate the predicted angular separation of the third component
to be 75~mas and its magnitude difference about 2~mag. Such values allow for the discovery of the
$3^{\mathrm{rd}}$ body using modern interferometric technique.

\subsection{DF Hya}

Another EB showing period changes is the system DF~Hya (AN~343.1934).%, GSC~00225-00913)
It is about 10.7~mag bright in $V$ filter, it has an orbital period about 0.3~days and belongs in
the W~UMa-type systems. It was discovered as a variable star by \cite{Hoffmeister1934}, who
classified the system as a short-period variable star. The most recent detailed analysis of its
light curve was published by \cite{1992DFHya}, who also derived its basic physical properties. Due
to the asymmetric shape of its light curve, both primary and secondary components were found to
have magnetic spots and the spectral type of the system was assumed to be G0V. The basic physical
parameters of the system derived by \cite{1992DFHya} are comparable with those derived by
\cite{Liu1990}.

\begin{figure}[b!]
 \plotone{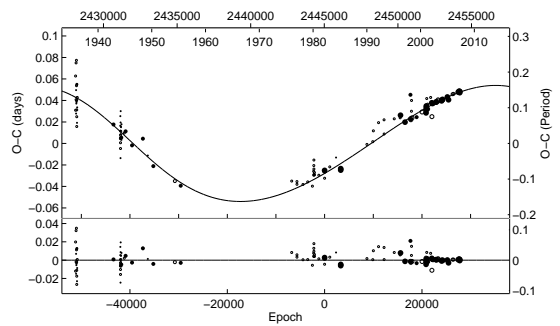}
 \caption{The $O-C$ diagram of DF~Hya (upper part), where the solid line represents the
 theoretical LITE variation caused by a 3$^\mathrm{rd}$ body and the $O-C$ residuals obtained
 after the subtraction of LITE (lower part). For the explanation of symbols see Fig.\ref{FigWZAnd1}.}
 \label{FigDFHya}
\end{figure}

As an explanation of the period changes in this system, all the previous period analyses proposed a
mass transfer between the two components \citep{Zhang1989IBVS}, or abrupt period jumps
\citep{Srivastava1991}. With the new set of up-to-date times of minima, counting altogether 143
data points, one is able to identify the long-term variation to be periodic instead of the steady
increase caused by a mass transfer. Especially, the new data points after the year 2000 evidently
deviate from the quadratic ephemeris (see Fig.\ref{FigDFHya}). The parameters of the LITE are given
in Table \ref{TableBig2}, the final fit is given in Fig.\ref{FigDFHya} and the predicted
third-body's minimal mass results in 0.84~\Mo. Assuming the component to be a main-sequence star,
then it should be of K1 spectral type and therefore a third light should be considered in the light
curve solution. Such a third light was not included in the light curve solution of
\cite{1992DFHya}. Using the same method as in the case of V803~Aql, one could also estimate the
value of the photometric distance of the additional component resulting in 190~pc, which yields the
predicted angular separation of 135~mas and a magnitude difference from the EB of about 1.3~mag.
Such a star could be easily detectable with the modern stellar interferometers.

\begin{table*}
\small \caption{The final results (part 2).} \label{TableBig2} \centering \scalebox{0.8}{
\begin{tabular}{c c c c }
\hline
 Parameter         &         PY~Lyr          &       FZ Ori            &         AH~Tau          \\
 \hline
 $JD_0$ $[$HJD$]$  & $2451663.563 \pm 0.005$ &$2450479.403 \pm 0.005$  &$2442750.359 \pm 0.002$   \\
 $P$ $[$day$]$     &$0.3857645 \pm 0.0000002$&$0.3999858 \pm 0.0000003$& $0.3326737 \pm 0.0000002$\\
 $p_3$ $[$yr$]$    &     $52.5 \pm 0.9$      &   $48.9 \pm 2.1$        &      $77.6 \pm 7.3$     \\
 $T_0$ $[$HJD$]$   &   $2451360 \pm 1780$    &  $2450500 \pm 460$      &   $2443600 \pm 1200$   \\
 $\omega$ $[$deg$]$&        $0 \pm 34$       &   $189.4 \pm 8.5$       &    $118.1 \pm 17.0$   \\
 $e$               &    $0.138 \pm 0.115$    &  $0.559 \pm 0.210$      &     $0.31 \pm 0.12$  \\
 $A$ $[$day$]$     &   $0.0395 \pm 0.0023$   & $0.0214 \pm 0.0010$     &   $0.0319 \pm 0.0023$\\
 $q$ $[$day$]$     &        --          & $(0.31 \pm 0.02)\cdot 10^{-10}$&           --       \\
 \hline
 $M_{12}$ [\Mo]    &         $2.5$           &       $2.1$            &        $1.61$        \\
 $d$ [pc]          &         $750$           &       $250$            &         281          \\
 $f(M_3)$ [\Mo]    &   $0.119 \pm 0.004$     &   $0.036 \pm 0.004$    &   $0.0289 \pm 0.0003$\\
 $M_{3,min}$ [\Mo] &    $1.17 \pm 0.04$      &    $0.65 \pm 0.02$     &    $0.51 \pm 0.01$  \\
 $a$ [mas]         &          $29$           &        $75$            &         83         \\
 \hline
\end{tabular}}
\end{table*}

\subsection{PY~Lyr}

The eclipsing binary system PY~Lyr (GSC~02136-03365) is of W~UMa-type and its magnitude is about
12.5 in $B$ filter. Both primary and secondary minima are about 0.6~mag deep and the orbital period
is about 0.4~d. Its spectral type was classified as F0 \citep{2006Malkov}, but it is only a
preliminary one.

\begin{figure}[b]
 \plotone{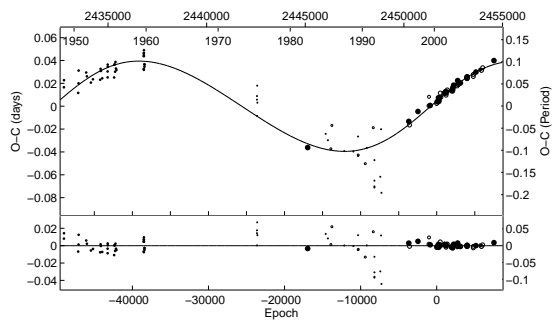}
 \caption{The $O-C$ diagram of PY Lyr (upper part), where the solid line represents the
 theoretical LITE variation caused by a 3$^\mathrm{rd}$ body and the $O-C$ residuals obtained
 after the subtraction of LITE (lower part). For the explanation of symbols see Fig.\ref{FigWZAnd1}.}
 \label{FigPYLyr}
\end{figure}

%The light curve of the system has not yet been analyzed, although it was observed by
%\cite{PYLyr2003}.
Precise CCD observations were carried out by \cite{PYLyr2006}, but they didn't include the third
light into their analysis. The radial velocities have not been measured, but many papers have been
published with times of minima. Information about a possible period change is given by
\cite{2003Pribulla}, who noticed some modulation of its orbital period, but they concluded
that this variation is uncertain. %v!: fast or complicated variation of the orbital period
On the other hand, \cite{Brat2001} published a paper on PY~Lyr, where the period changes were
described by two period jumps -- near 1967 and 1987.

Collecting all the minima times, one gets a set of 123 data points. Fig.\ref{FigPYLyr} represents
the $O-C$ diagram of all these measurements, where the period variation is clearly visible. A
period of about 50~years is now well covered and the resulting parameters of the predicted LITE
variation are given in Table \ref{TableBig1}. Assuming that the mass of the eclipsing pair is about
2.5~\Mo~(according to its spectral type), the minimal mass of the predicted third component results
in 1.17~\Mo, which is approximately the same value of mass as the primary and secondary component.
Therefore, the third light can be easily detectable in the light curve solution. Regrettably, the
parallax and the distance to this system is not known, but it could be estimated using the same
method as in the previous case. The value of the system's photometric distance results in 750~pc,
therefore the predicted angular separation of the third component is about 29~mas and its magnitude
difference from the EB is about 1.3~mag. Such a component would be hardly observable
interferometrically.

\begin{table*}
\small \caption{The light curve parameters of PY~Lyr. The '*' mark indicates the assumed value.} \label{Table3}
\centering \scalebox{0.85}{
\begin{tabular}{c c|| c c c c c c c c}
\hline
  $T_1$[K] &6980$^*$&     &$L_1$[\%]&$L_2$[\%]&$L_3$[\%]& $x_1$ & $x_2$ \\
  $T_2$[K] &  7042  & $B$ &  47.0   &  34.3   &  18.7   & 0.607 & 0.603 \\
  $i$[deg] & 80.379 & $V$ &  46.7   &  33.8   &  19.5   & 0.498 & 0.495 \\
  $q$      & 0.6596 & $R$ &  46.7   &  33.6   &  19.7   & 0.417 & 0.414 \\
 $\Omega_1$& 3.018  & $I$ &  46.1   &  33.0   &  20.9   & 0.339 & 0.335 \\ \hline
 \multicolumn{2}{c}{Spot parameters:} & Latitude[deg]& Longitude[deg] & Radius[deg] & $T_{spot}/T_{surface}$\\
 \multicolumn{2}{c}{}&71.9 & 278.3  & 18.6    &  0.798 \\
 \hline
\end{tabular}}
\end{table*}

\begin{figure}[b]
 \plotone{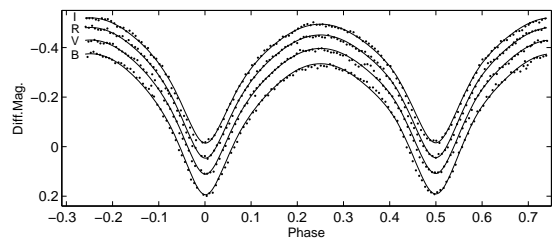}
 \caption{The light curves of PY~Lyr in $B$, $V$, $R$, and $I$ filters, respectively. The solid lines represent
 the solution according to the parameters given in Table \ref{Table3}.}
 \label{FigPYLyrLC}
\end{figure}

The photometric data obtained by one of us (V.M.), we reanalysed again by taking into consideration
the third body's contribution to the total light of the system. The software PHOEBE 0.29d, which is
based on the Wilson-Devinney code, was used in order to extract the new model of the system. $B$,
$V$, $R$, and $I$ observations of this system were analyzed (see Fig.\ref{FigPYLyrLC}), yielding a
new set of physical parameters given in Table \ref{Table3}, where $T_i$, $L_i$, $\Omega_i$, and
$x_i$ denote the temperature, the luminosity, the modified Kopal potential, and the limb-darkening
coefficients for primary and secondary, respectively. The "mode 3" was used for computing (hence
$\Omega_1$ = $\Omega_2$) and the eccentricity was set to 0 (circular orbit). The value of
temperature of the primary component was assumed from its spectral type. The limb-darkening
coefficients were interpolated from van~Hamme's tables (see \citealt{vanHamme1993}). The values of
gravity brightening and bolometric albedo coefficients were set at their suggested values for
convective atmospheres (see \citealt{Lucy1968}), i.e. $g_1 = g_2 = 0.32$, $A_1 = A_2 = 0.5$. Also
the synchronous rotation was assumed for each star ($F_1 = F_2 = 1.0$). A spot in the primary
component has been used due to the presence of the O'Connell effect. The contribution of the third
light to the total luminosity of the system is significant. Its value from the light curve solution
results in approximately (19.7 $\pm$5)\%. On the other hand, the value predicted according to the
LITE variation is about (23 $\pm$ 5)\%, so it is in very good agreement with each other.

\subsection{FZ Ori}

The EB system FZ Ori (HD~288166) has been discovered to be a variable by \cite{Hoffmeister1934}. It
is a W~UMa-type system with an orbital period of about 0.4~days and a brightness of about 10.8~mag
in $V$ filter. Its spectral type was estimated as G0 \citep{KholopovGCVS}.

\begin{figure}[b]
 \plotone{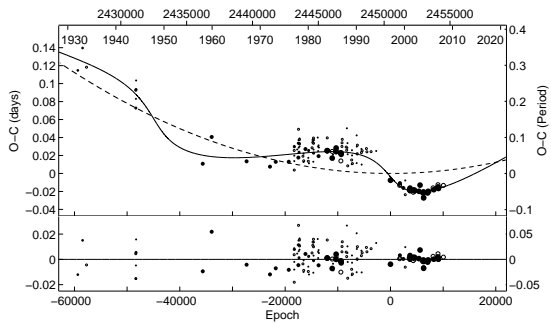}
 \caption{The $O-C$ diagram of FZ~Ori, (upper part), where the solid line represents the
 theoretical LITE variation caused by a 3$^\mathrm{rd}$ body and the dashed line represents the
 quadratic ephemeris, and the $O-C$ residuals obtained after the subtraction of LITE (lower part).
 For the explanation of symbols see Fig.\ref{FigWZAnd1}.}
 \label{FigFZOri}
\end{figure}

The light curve of the star was analyzed a few times in the past, but no one of the analyses has
been very detailed. One of them was published by \cite{Rukmini2001} and the most recent one by
\cite{Byboth2004}. Both of these studies indicate a slightly asymmetric light curve, which was
explained in the latter paper by the presence of a spot on the primary component. A period analysis
was performed by \cite{Alawy1993}, who also mentioned a possible cyclic changing of the period, but
no satisfactory solution was presented.

Our new period analysis is based on a much larger data set, containing 153 times of minima. From
these measurements it is possible to identify the steady period increase besides the cyclic
variation of its orbital period with a periodicity of about 50~yrs (see Fig.\ref{FigFZOri} for the
$O-C$ fit and Table \ref{TableBig2} for the final parameters). According to the value of the
quadratic-term coefficient, one gets surprisingly high value of conservative mass transfer rate
between the components, of about $5.1 \cdot 10^{-7}$~\Mo/yr. A similar result is also presented in
\cite{Rukmini2001}, where the authors deduce that the system is now in the stage of mass transfer.
Assuming that the masses of the primary and secondary components are $M_1 = 1.1$~\Mo~ and $M_2 =
1.0$~\Mo~ (according to \citealt{Hec1988}) respectively, minimal mass of the third body results in
0.65~\Mo. Such a component could be probably detectable in the light curve solution and also could
be evident in the spectrum of the system, but no such attempt has been carried out so far. Its
photometric distance results in approximately 250~pc, and therefore the predicted angular
separation is about 75~mas, while the magnitude difference from the EB is about 2.8~mag. A
detection of such a component is hence near the limits of the current interferometric techniques.

\subsection{AH Tau}

AH Tau is an eclipsing binary of W~UMa-type, its apparent brightness is about 11.4~mag in $V$
filter, its orbital period is about 0.33~days and the spectral type was classified as G1p
\citep{Brancewicz1980}. The first photographic light curve was observed and briefly analyzed by
\cite{Binnendijk1950}. The most detailed analysis of its light curve was presented in
\cite{Liu1991}, resulting in a set of basic physical parameters.

\begin{figure}[t]
 \plotone{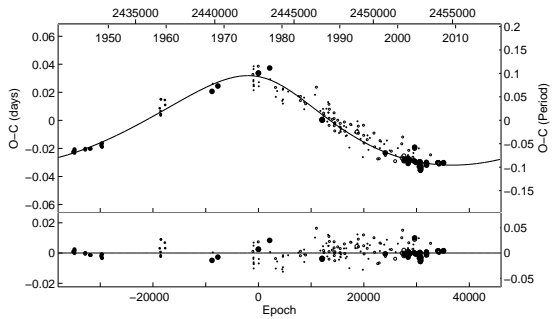}
 \caption{The $O-C$ diagram of AH~Tau (upper part), where the solid line represents the
 theoretical LITE variation caused by a 3$^\mathrm{rd}$ body and the $O-C$ residuals obtained
 after the subtraction of LITE (lower part). For the explanation of symbols see Fig.\ref{FigWZAnd1}.}
 \label{FigAHTau}
\end{figure}

A period analysis was performed by \cite{Yang2002}, who also noticed variable period changes and
they concluded that their existence was due to the magnetic activity, the mass loss and the radius
swelling of the components. Our new data set consists of more data points (204 times of minima).
The final $O-C$ diagram fitted by the theoretical curves is displayed in Fig.\ref{FigAHTau} and the
final LITE parameters are given in Table \ref{TableBig2}. The parabolic curve due to the steady
period decrease could not fit very well the most recent data points, but LITE hypothesis is able to
describe also the behavior of the newest observations.

The LITE parameters lead to the predicted value of the third mass, resulting in 0.51~\Mo. Contribution of such a
body could be observable in a precisely measured light curve of this system. \cite{Liu1991} also derived a
photometric parallax of this system, resulting in a value of 281~pc, which yields the predicted angular distance
of such body of about 83~mas, which is within the limits for the modern stellar interferometers. On the other
hand, the magnitude difference between the third component and the eclipsing pair is about 3.7~mag, which is too
high to expect such a body to be discovered.

\subsection{Alternative explanation of the orbital period changes} \label{AlternativeExpl}

The O-C residuals show additional non-periodic variations in the $O-C$ diagrams, which cannot be
described by applying only the LITE hypothesis. The residuals in Figs. \ref{FigWZAnd1} to
\ref{FigAHTau} show also additional variations with variable periods and amplitudes much lower than
that of LITE. These could be caused by the presence of stellar convection zones, in an agreement
with the so-called Applegate's mechanism, see e.g. \cite{Applegate1992}, \cite{Lanza1998}, or
\cite{Hoffman2006}. Such an effect could play a role, because the spectral types of many of the
components are later than F5 (see \cite{Zavala2002} for a detailed analysis). To conclude, for a
better description of the observed period variations of these systems, the magnetic activity cycles
could be present together with the LITE. On the other hand, one has to take into consideration that
the spectral types of most of these binaries were not derived from their spectra, but only on the
basis of their photometric indices and therefore are not very reliable.

The only system, where one could estimate the variation of the quadruple moment (see
\citealt{Applegate1992}) required to explain the long-term period variations, is AH~Tau, where the
semi-major axis of the orbit from the light curve solution is known. Using the following equation
$$ \Delta P = A \sqrt{2 (1 - \cos \{ 2 \pi P/p_3 \} )} $$ \citep{Rovit2000} one could compute the
amplitude of the period oscillation. The period variation $\Delta P / P$ can be used for
calculating the variation of the quadruple moment $\Delta Q$, using the equation \citep{Lanza2002}
$$ \frac{\Delta P}{P} = -9 \frac{\Delta Q}{Ma^2}.$$ This quantity results in $\Delta Q = (6.20 \pm 0.60)
\cdot 10^{49}$~g$\cdot$cm$^2$, which is not inside the limits for active binaries (range of values
from $10^{50}$ to $10^{51}$~g$\cdot$cm$^2$) and therefore the variation in AH~Tau could not be
explained by this mechanism.

\section{Discussion and conclusions}

Six contact eclipsing binaries were analyzed for the presence of
LITE on the basis of their $O-C$ diagram analysis and the
times-of-minima variations. A few new observations of these
systems were obtained and used in the present analysis. All of the
studied systems show apparent changes of their orbital periods,
which could be explained as a result of a third component orbiting
the EB around their common center of mass.

Such a variation has usually a period of the order of decades, as one can see from
Figs.\ref{FigWZAnd1}--\ref{FigAHTau}, which can be described by applying the LITE hypothesis sufficiently. In
the case of FZ~Ori the quadratic term in the light elements was also used. This could be explained as a mass
transfer between the two components, which is a common procedure in contact systems. The conservative
mass-transfer rate was calculated.

Regrettably, in most of the above systems no detailed analysis
(neither photometric nor spectroscopic) has been made so far. The
spectral types and the masses of the individual components in the
systems are only approximate, so the parameters of the predicted
third bodies are also affected by relatively large errors. Due to
missing information about the distances to these binaries, we have
used a photometric parallax for the distance determination and
therefore also the predicted angular separations of the third
components could be estimated. It is obvious that only further
detailed photometric, as well as spectroscopic and interferometric
analysis would reveal the nature of these systems and confirm or
reject the third-body hypothesis.

\section{Acknowledgments}

This investigation was supported for P.Z. \& M.W. by the Czech-Greek project of collaboration No.
7-2006-5 {\it Photometry and Spectroscopy of Binaries} of Ministry of Education, Youth and Sport of
the Czech Republic and by the Czech Science Foundation grant No. 205/06/0217, and for A.L., P.N.,
V.M. \& K.G. by the Special Account for Research Grants 70/3/8680 of the National \& Kapodistrian
University of Athens, Greece. This research has made use of the SIMBAD database, operated at CDS,
Strasbourg, France, and of NASA's Astrophysics Data System Bibliographic Services.


\begin{thebibliography}{}
 \bibitem[Alawy(1993)]{Alawy1993} Alawy, A.~A.~E.-B.\ 1993, \apss, 207, 171
 \bibitem[Albayrak et al.(1999)]{Albayrak1999} Albayrak, B., Fikri {\"O}zeren, F., Ekmek{\c c}i, F., \& Demircan, O.\ 1999, RevMexAA, 35, 3
 \bibitem[Applegate(1992)]{Applegate1992} Applegate, J.~H.\ 1992, \apj, 385, 621
 \bibitem[Binnendijk(1950)]{Binnendijk1950} Binnendijk, L.\ 1950, \bain, 11, 209
 \bibitem[Borkovits \& Heged\"us(1996)]{BorkovitsHegedus} Borkovits, T., \& Heged\"us, T.\ 1996, \aaps, 120, 63
 \bibitem[Brancewicz \& Dworak(1980)]{Brancewicz1980} Brancewicz, H.~K., \& Dworak, T.~Z.\ 1980, Acta Astronomica, 30, 501
 \bibitem[Br\'at (2001)]{Brat2001} Br\'at, L.\ 2001, Perseus, 1, 8
 \bibitem[Byboth et al.(2004)]{Byboth2004} Byboth, K.~N., Markworth, N.~L., \& Bruton, W.~B.\ 2004, Informational Bulletin on Variable Stars, 5554, 1
 \bibitem[Cook(1948)]{1948Cook} Cook, A.~F.\ 1948, \aj, 53, 211
% \bibitem[Demircan et al.(2003)]{PYLyr2003} Demircan, O., et al.\ 2003, Informational Bulletin on Variable Stars, 5364, 1
 \bibitem[Frieboes-Conde \& Herczeg(1973)]{FCH73} Frieboes-Conde, H., \& Herczeg, T.\ 1973, \aaps, 12, 1
 \bibitem[Hall(1989)]{Hall1989} Hall, D.~S.\ 1989, Space Science Reviews, 50, 219
 \bibitem[Harmanec(1988)]{Hec1988} Harmanec, P.\ 1988, Bulletin of the Astronomical Institutes of Czechoslovakia, 39, 329
 \bibitem[Hoffman et al.(2006)]{Hoffman2006} Hoffman, D.~I., Harrison, T.~E., McNamara, B.~J., Vestrand, W.~T., Holtzman, J.~A., \& Barker, T.\ 2006, \aj, 132, 2260
 \bibitem[Hoffmeister(1934)]{Hoffmeister1934} Hoffmeister, C.\ 1934, Astronomische Nachrichten, 253, 195
 \bibitem[Irwin(1959)]{Irwin1959} Irwin, J.~B.\ 1959, \aj, 64, 149
 \bibitem[Kholopov(1985)]{KholopovGCVS} Kholopov, P.N., Samus, N.N., Frolov, M.S., Goranskij, V.P.,
    Gorynya, N.A., Kireeva, N.N., Kukarkina, N.P., Kurochkin, N.E., Medvedeva, G.I., Perova, N.B., and Shugarov, S.
    Yu., General Catalogue of Variable Stars, 4th Edition, Volumes I-III
 \bibitem[Kreiner et al.(2001)]{Kreiner2001} Kreiner, J.~M., Kim, C.-H., \& Nha, I.-S.\ 2001, An Atlas of O-C Diagrams of Eclipsing Binary Stars / by Jerzy M.~Kreiner, Chun-Hwey Kim, Il-Seong Nha.~Cracow, Poland: Wydawnictwo Naukowe Akademii Pedagogicznej.~2001.
 \bibitem[Kukarkin et al.(1971)]{1971GCVS} Kukarkin, B.~V., Kholopov, P.~N., Pskovsky, Y.~P., Efremov, Y.~N., Kukarkina, N.~P., Kurochkin, N.~E., \& Medvedeva, G.~I.\ 1971, General Catalogue of Variable Stars, 3rd ed.~(1971), 0
 \bibitem[Kwee \& van Woerden(1956)]{Kwee} Kwee, K.~K., \& van Woerden, H.\ 1956, \bain, 12, 327
 \bibitem[Lanza et al.(1998)]{Lanza1998} Lanza, A.~F., Catalano, S., Cutispoto, G., Pagano, I., \& Rodono, M.\ 1998, \aap, 332, 541
 \bibitem[Lanza \& Rodon{\`o}(2002)]{Lanza2002} Lanza, A.~F., \& Rodon{\`o}, M.\ 2002, Astronomische Nachrichten, 323, 424
 \bibitem[Liu et al.(1990)]{Liu1990} Liu, Q.-Y., Yang, Y.-L., Zhang, Y.-L., Wang, B., \& Zhang, Z.-S.\ 1990, Acta Astronomica Sinica, 31, 237
 \bibitem[Liu et al.(1991)]{Liu1991} Liu, Q.-Y., Yang, Y.-L., Zhang, Y.-L., \& Wang, B.\ 1991, Acta Astrophysica Sinica, 11, 143
 \bibitem[Lucy(1968)]{Lucy1968} Lucy, L.~B.\ 1968, \apj, 151, 1123
 \bibitem[Malkov et al.(2006)]{2006Malkov} Malkov, O.~Y., Oblak, E., Snegireva, E.~A., \& Torra, J.\ 2006, \aap, 446, 785
 \bibitem[Manimanis et al.(2006)]{PYLyr2006} Manimanis, V.~N., Niarchos, P.~G., \& Gazeas, K.~D.\ 2006, Recent Advances in Astronomy and Astrophysics, 848, 417
 \bibitem[Mayer(1990)]{Mayer1990} Mayer, P.\ 1990, Bulletin of the Astronomical Institutes of Czechoslovakia, 41, 231
 \bibitem[Niarchos et al.(1992)]{1992DFHya} Niarchos, M., Hoffmann, M., \& Duerbeck, H.~W.\ 1992, \aap, 258, 323
 \bibitem[Oh(1991)]{1991PASA} Oh, K.-D.\ 1991, Proceedings of the Astronomical Society of Australia, 9, 289
 \bibitem[Pribulla et al.(2003)]{2003Pribulla} Pribulla, T., Kreiner, J.~M., \& Tremko, J.\ 2003, Contributions of the Astronomical Observatory Skalnate Pleso, 33, 38
 \bibitem[Rafert(1982)]{1982PASP} Rafert, J.~B.\ 1982, \pasp, 94, 485
 \bibitem[Rovithis-Livaniou et al.(2000)]{Rovit2000} Rovithis-Livaniou, H., Kranidiotis, A.~N., Rovithis, P., \& Athanassiades, G.\ 2000, \aap, 354, 904
 \bibitem[Rukmini et al.(2001)]{Rukmini2001} Rukmini, J., Vivekananda Rao, P., \& Ausekar, B.~D.\ 2001, Bulletin of the Astronomical Society of India, 29, 323
 \bibitem[Samec et al.(1993)]{Samec1993PASP} Samec, R.~G., Su, W., \& Dewitt, J.~R.\ 1993, \pasp, 105, 1441
 \bibitem[Shapley(1923)]{Leavitt1923} Shapley, H.\ 1923, Harvard College Observatory Bulletin, 790, 1
 \bibitem[Shaw(1994)]{1994Shaw} Shaw, J.~S.\ 1994, Memorie della Societa Astronomica Italiana, 65, 95
 \bibitem[Srivastava(1991)]{Srivastava1991} Srivastava, R.~K.\ 1991, \apss, 181, 15
 \bibitem[Sterken(2005)]{Sterken2005} Sterken, C.\ 2005, The Light-Time Effect in Astrophysics: Causes and cures of the O-C diagram, 335
 \bibitem[Svechnikov \& Kuznetsova(1990)]{Svechnikov1990} Svechnikov, M.~A., \& Kuznetsova, E.~F.\ 1990, Sverdlovsk : Izd-vo Ural'skogo universiteta, 1990
 \bibitem[van Hamme(1993)]{vanHamme1993} van Hamme, W.\ 1993, \aj, 106, 2096
 \bibitem[Wolf et al.(2004)]{Wolf2004} Wolf, M., Mayer, P., Zasche, P., \v{S}arounov\'a, L., \& Zejda, M.\ 2004, Spectroscopically and Spatially Resolving the Components of the Close Binary Stars, 318, 255
 \bibitem[Woltjer(1922)]{Woltjer1922} Woltjer, J., Jr.\ 1922, \bain, 1, 93
 \bibitem[Yang \& Liu(2002)]{Yang2002} Yang, Y., \& Liu, Q.\ 2002, \aap, 390, 555
 \bibitem[Zasche et al.(2008)]{ZLWN2008} Zasche, P., Liakos, A., Wolf, M., \& Niarchos, P.\ 2008, New Astronomy, 13, 405
 \bibitem[Zavala et al.(2002)]{Zavala2002} Zavala, R.~T., et al.\ 2002, \aj, 123, 450
 \bibitem[(1925)]{1925AN} Zessewitsch, W.\ 1925, Astronomische Nachrichten, 223, 149
 \bibitem[Zhang et al.(1989)]{Zhang1989IBVS} Zhang, Y., Liu, Q., Yang, Y., Wang, B., \& Zhang, Z.\ 1989, Informational Bulletin on Variable Stars, 3349, 1
 \bibitem[Zhang \& Zhang(2006)]{WZAnd2006NewA} Zhang, X.~B., \& Zhang, R.~X.\ 2006, New Astronomy, 11, 339
\end{thebibliography}
\end{document}